\documentclass[reprint,aps,pra,superscriptaddress]{revtex4-1}

\usepackage{braket,mathtools,amssymb}
\usepackage{color}

\begin{document}

\title{Feasibility of efficient room-temperature solid-state sources of indistinguishable single photons using ultrasmall mode volume cavities}

\author{Stephen Wein}
\author{Nikolai Lauk}
\author{Roohollah Ghobadi}
\author{Christoph Simon}
\affiliation{Institute for Quantum Science and Technology and
Department of Physics and Astronomy, University of Calgary, Calgary, Alberta,
Canada T2N 1N4}

%\affiliation[*]{Corresponding author: scwein@ucalgary.ca}

\begin{abstract}
Highly efficient sources of indistinguishable single photons that can operate at room temperature would be very beneficial for many applications in quantum technology. We show that the implementation of such sources is a realistic goal using solid-state emitters and ultrasmall mode volume cavities. We derive and analyze an expression for photon indistinguishability that accounts for relevant detrimental effects, such as plasmon-induced quenching and pure-dephasing. We then provide the general cavity and emitter conditions required to achieve efficient indistinguishable photon emission, and also discuss constraints due to phonon sideband emission. Using these conditions, we propose that a nanodiamond negatively charged silicon-vacancy center combined with a plasmonic-Fabry-P\'{e}rot hybrid cavity is an excellent candidate system.
\end{abstract}
\date{\today}

\maketitle

\section{Introduction}
While there is substantial excitement about quantum technology and quantum information processing (QIP), many practical applications are still held back by the fact that critical components are restricted to operating at cryogenic temperatures. Trying to overcome the thermal restrictions of quantum devices also tests fundamental questions about the physical regimes in which quantum processes can exist and be manipulated.

Indistinguishable single-photon sources (SPSs) are basic components of numerous different optical QIP implementations. They are required for tasks such as linear-optical quantum computing \cite{kok2007} and boson sampling \cite{spring2012, broome2013}. In addition, an efficient indistinguishable photon source can be used to construct quantum repeaters and would assist in the development of a quantum internet \cite{gisin2002,sangouard2007, kimble2008, simon2017}.

The most common way to generate indistinguishable photons at room temperature is by heralded spontaneous parametric down-conversion (SPDC) \cite{kwiat1995}. This technique has seen pioneering success in quantum research, yet its probabilistic nature limits its range of applications, although this limitation could in principle be addressed by multiplexing many SPDC sources \cite{joshi2017,kaneda2018}.

In contrast, individual quantum emitters not only promise near-deterministic single-photon emission at room temperature, but moreover the emitted photon could be entangled with coherent solid-state spins \cite{balasubramanian2009}. This would allow many QIP applications to be performed at room temperature, including optically-mediated entanglement of distant spins in solids \cite{bernien2013}.

Low-temperature indistinguishable SPSs have been achieved \cite{santori2002} and are becoming more efficient \cite{ding2016, somaschi2016}. Ref. \cite{grange2015} showed that relatively inefficient indistinguishable SPSs can be realized beyond the low-temperature regime using weakly-coupled narrow-bandwidth micro-cavities. These results were applied to extend quantum dot indistinguishable SPS operation up to 20 K \cite{grange2017}. Distinguishable SPSs have also been demonstrated at room temperature using solid-state quantum emitters \cite{aharonovich2016}. In particular, recently, plasmonic cavities have been proposed to enhance emission rates for distinguishable SPSs in spite of high losses \cite{bozhevolnyi2016,bozhevolnyi2017}.

Achieving indistinguishable photon emission from a solid-state emitter is a difficult task, especially at higher temperatures. The main problem is that optical transitions in solid-state materials experience rapid phonon-induced dephasing that homogeneously broadens the zero-phonon line (ZPL) at room temperature \cite{davies1974, fu2009}. This dephasing reduces the degree of indistinguishability between emitted photons \cite{kiraz2004, grange2015}. In addition, phonon-assisted optical transitions can produce a phonon sideband (PSB) that reduces indistinguishability and consequently must be filtered, which sacrifices efficiency. If the PSB spectrum overlaps with the desired ZPL emission, it cannot be entirely filtered and hence fundamentally limits indistinguishability \cite{smith2017b, smith2017}. Furthermore, when using plasmonic materials to enhance emission, plasmon-induced quenching poses another detrimental effect that affects both efficiency and indistinguishability.  A very recent analysis on a single spherical metallic nanoparticle with the goal of producing an on-chip room temperature single-photon source \cite{peyskens2017} found that simultaneous high efficiency and high indistinguishability are difficult to achieve with integrated plasmonics.

Here we show that achieving high efficiency and high indistinguishability simultaneously at room temperature should be possible by using ultrasmall mode volume cavities and by operating near the boundary between the strong coupling and bad-cavity regimes (to be defined precisely below). We argue that such systems are within reach of current technology, and we
provide theoretical guidelines for designing successful plasmonic cavities for this purpose.

A common approach to improving indistinguishability and efficiency simultaneously is to place a quantum emitter inside a cavity.  In addition to suppressing off-resonant PSB emission, the cavity reduces the emitter lifetime through the Purcell effect \cite{purcell1946} and allows the photon to be emitted before the optical coherence is destroyed by interactions with the phonon bath. At room temperature, this fast optical dephasing is very difficult to overcome, requiring Purcell factors exceeding $10^4$ for most emitters \cite{grange2015}. Obtaining a large Purcell factor can be accomplished by either increasing the cavity quality factor or by decreasing the effective mode volume. However, for highly dissipative emitters, increasing the quality factor too high prolongs the interaction between the cavity photon and the dissipative emitter, which causes optical dephasing. As a result, it is necessary to use cavities with mode volumes far below the diffraction limit.

Such ultrasmall mode volume cavities have seen significant development over the last decade and, in particular, the last few years \cite{robinson2005,seo2009,kuttge2009,russell2010,kang2011,hu2016,choi2017,gurlek2017,santhosh2016,chikkaraddy2016,peng2017}. Many of these cavities utilize plasmonic materials to concentrate the electromagnetic field along a material interface in the form of plasmon-polaritons \cite{seo2009,kuttge2009,russell2010,kang2011,santhosh2016,chikkaraddy2016,gurlek2017,peng2017}. There are also interesting proposals for pure-dielectric ultrasmall cavities \cite{robinson2005,hu2016,choi2017} and plasmonic-Fabry-P\'{e}rot hybrid cavities \cite{gurlek2017,peng2017}. However, the application of ultrasmall mode volume cavities to quantum information processing is still relatively unexplored.

\section{System and figures of merit}

We now describe our proposed approach in detail. We begin with the interaction Hamiltonian for a driven two-level system coupled to a resonant cavity:
\begin{equation}
\label{hamiltonian}
\frac{1}{\hbar}\hat{H} = g(\hat{\sigma}_-\hat{a}^\dagger+\hat{\sigma}_+\hat{a})+\frac{\Omega}{2}(\hat{\sigma}_-+\hat{\sigma}_+),
\end{equation}
where $\hat{\sigma}_-$ ($\hat{\sigma}_+$) is the two-level system lowering (raising) operator, $\hat{a}$ ($\hat{a}^\dagger$) is the cavity mode annihilation (creation) operator, $g$ is the cavity coupling factor, and $\Omega$ is the driving Rabi frequency. The driving term of the Hamiltonian significantly complicates the derivation for indistinguishability. However, the single-photon purity of emission from a two-level system depends on how the system is excited. Slow excitation allows for multi-photon emission, which increases the second-order intensity correlation at zero time delay ($g^{(2)}(0)$). For a high single-photon purity ($g^{(2)}(0)\simeq 0$), excitation of the emitter will require ultrafast optical control, which has been demonstrated for defects in diamond \cite{bassett2014,becker2016}. The common practice, which we also adopt, is to assume $\Omega\gg g$ during excitation so that the system is effectively instantaneously prepared in the excited state \cite{grange2015,peyskens2017}, implying $g^{(2)}(0)\simeq 0$. We then explore the emission dynamics as governed by the system when $\Omega=0$. This allows us to make use of the single-excitation approximation and decouple the optical Bloch equations.

The interaction between a quantum emitter and the phonon bath will broaden the emitter's ZPL and could also create a PSB. Recently, the effect of the PSB on the indistinguishability of single-photon sources was studied \cite{smith2017,smith2017b}. It was found that the indistinguishability is limited by the fraction of PSB not filtered by the cavity. This limitation can be highly detrimental at room temperature where the ZPL is very broad, which necessitates a broad cavity that might not filter the PSB. Therefore, an ideal emitter for an efficient room-temperature indistinguishable SPS should have a small PSB that is spectrally well-separated from its ZPL. For such an emitter, pure-dephasing and possible quenching effects will be the primary limitation. Consequently, we first consider a Markovian system that neglects non-Markovian PSB effects in order to describe the most ideal parameter regime for a good (Markovian) emitter. Then we estimate the correction induced by a non-zero PSB using the results of Iles-Smith \emph{et al.} \cite{smith2017}.

We described the dissipative dynamics of the system using the Markovian master equation:
\begin{equation}
\label{mastereq}
\dot{\rho} = -\frac{i}{\hbar}[\hat{H},\rho] + \kappa\mathcal{D}(\hat{a})\rho + \gamma^\star\mathcal{D}(\hat{\sigma}_+\hat{\sigma}_-)\rho+\gamma\mathcal{D}(\hat{\sigma}_-)\rho,
\end{equation}
with $\mathcal{D}(\hat{A})\rho = \hat{A}\rho\hat{A}^\dagger - \{\hat{A}^\dagger\hat{A},\rho\}/2$, bare cavity linewidth $\kappa$, emitter lifetime $1/\gamma$, and pure-dephasing rate $\gamma^\star$. The cavity linewidth $\kappa=\kappa_{r}+\kappa_\text{nr}$ has a radiative part $\kappa_\text{r}$ and non-radiative part $\kappa_\text{nr}$. The cavity quality factor is $Q = \omega/\kappa$ and we define the bare cavity quantum efficiency as $\eta_\text{r}=\kappa_\text{r}/\kappa$. Similarly, the natural decay rate $\gamma=\gamma_\text{r}+\gamma_\text{nr}$ also has a radiative part $\gamma_\text{r}$ and a non-radiative part $\gamma_\text{nr}$.

We can state the requirements for high efficiency and indistinguishability in terms of the parameters introduced above. First, the rate of population transfer from the emitter to the cavity must exceed the optical dephasing rate: $R>\gamma^\star$, where $R=4g^2/\kappa$ is related to the standard Purcell factor expression \mbox{$P=(3/4\pi^2)(\lambda/n)^3(Q/V)$} by $R=\gamma_\text{r}P$. Here, $V$ is the effective mode volume of a cavity with resonance wavelength $\lambda$ and refractive index $n$. For a broad emitter, $P=R/\gamma_\text{r}$ is not the effective Purcell enhancement; rather, it serves as a convenient cavity metric that is also useful when defining regime boundaries. %See the appendix for a discussion regarding the effective Purcell enhancement. 
Second, the photon must escape the cavity before it is dephased by the emitter: $\kappa>\gamma^\star$. These conditions define a region that overlaps with the strong-coupling regime ($2g>\gamma+\kappa+\gamma^\star$) and the bad-cavity regime ($2g<\gamma+\kappa+\gamma^\star$ and $\kappa>\gamma^\star$). We will call this region of high efficiency and indistinguishability the critical regime to distinguish it from other regimes it overlaps.

To quantify the efficiency of the system, we use the cavity efficiency $\beta = \kappa\int_0^\infty \braket{\hat{a}^\dagger(t)\hat{a}(t)}\text{d}t$ \cite{grange2015}. The derivation is given in the Appendix and the result is:
\begin{equation}
\label{effdef}
\beta=\frac{R\kappa}{R(\gamma+\kappa)+\gamma(\gamma+\gamma^\star+\kappa)},
\end{equation}
In our analysis, we only explicitly compute $\beta$---the efficiency of population dissipated through the cavity mode---hereafter referred to as the intrinsic cavity efficiency. The total radiative quantum efficiency is given by $\beta\eta_\text{r}$. The quantity $\beta\eta_\text{r}$ still does not account for photon collection efficiency, which will also affect the total efficiency of the SPS. We discuss the radiative quantum efficiency and photon collection efficiency more in section \ref{sec:candidatePSB}. The efficiency of an SPS in the presence of plasmonic materials has been studied \cite{bozhevolnyi2016,bozhevolnyi2017}. However, for many QIP applications, the efficiency is not the only relevant metric, but the indistinguishability of the emitted photons is also essential. 

The metric that we use for indistinguishability is derived from the probability that two photons emitted from the same source interfere and bunch at a beamsplitter \cite{kiraz2004, grange2015}:
\begin{equation}
\label{inddef}
I = \frac{\int_0^\infty\int_0^\infty|\braket{\hat{a}^\dagger(t+\tau)\hat{a}(t)}|^2\text{d}t\text{d}\tau}{\int_0^\infty\int_0^\infty\braket{\hat{a}^\dagger(t+\tau)\hat{a}(t+\tau)}\braket{\hat{a}^\dagger(t)\hat{a}(t)}\text{d}t\text{d}\tau}.
\end{equation}

An ideal indistinguishable SPS should have near-unity indistinguishability $I$ and efficiency $\beta$. To this end, we focus on maximizing the indistinguishability-efficiency product $I\beta$.

\section{Indistinguishability and quenching}

Using the quantum regression theorem \cite{gardiner2004}, we derived an expression for indistinguishability valid in the critical regime for arbitrary $\gamma$. As we will show, this allows the expression to capture possible effects of plasmon quenching. The details of the derivation and the full solution valid for arbitrary $\gamma$ are given in the Appendix. Here we only write the expression in the case that $\gamma<\gamma^\star<\kappa$, which arises when quenching is weak:
\begin{equation}
\label{ind}
I=\frac{R^2\kappa^2\left(1+I_1\right)}{(R+\gamma)(\kappa+\gamma)(R+\gamma+\gamma^\star)(\kappa+\gamma+\gamma^\star)\beta^2},
\end{equation}
where $I_1=(\gamma^\star/\kappa)(6\kappa-R)/(3\kappa+4R)$ and $\beta$ is given by Eq. (\ref{effdef}). This expression is accurate in the critical regime to first order in $\gamma^\star/\kappa$.

\begin{figure}[t]
\centering
\begin{flushleft}\mbox{\hspace{4mm}(a)\hspace{36mm}(c)}\end{flushleft}\vspace{-5mm}
\includegraphics[trim={5mm 5mm 0 0},scale=0.405]{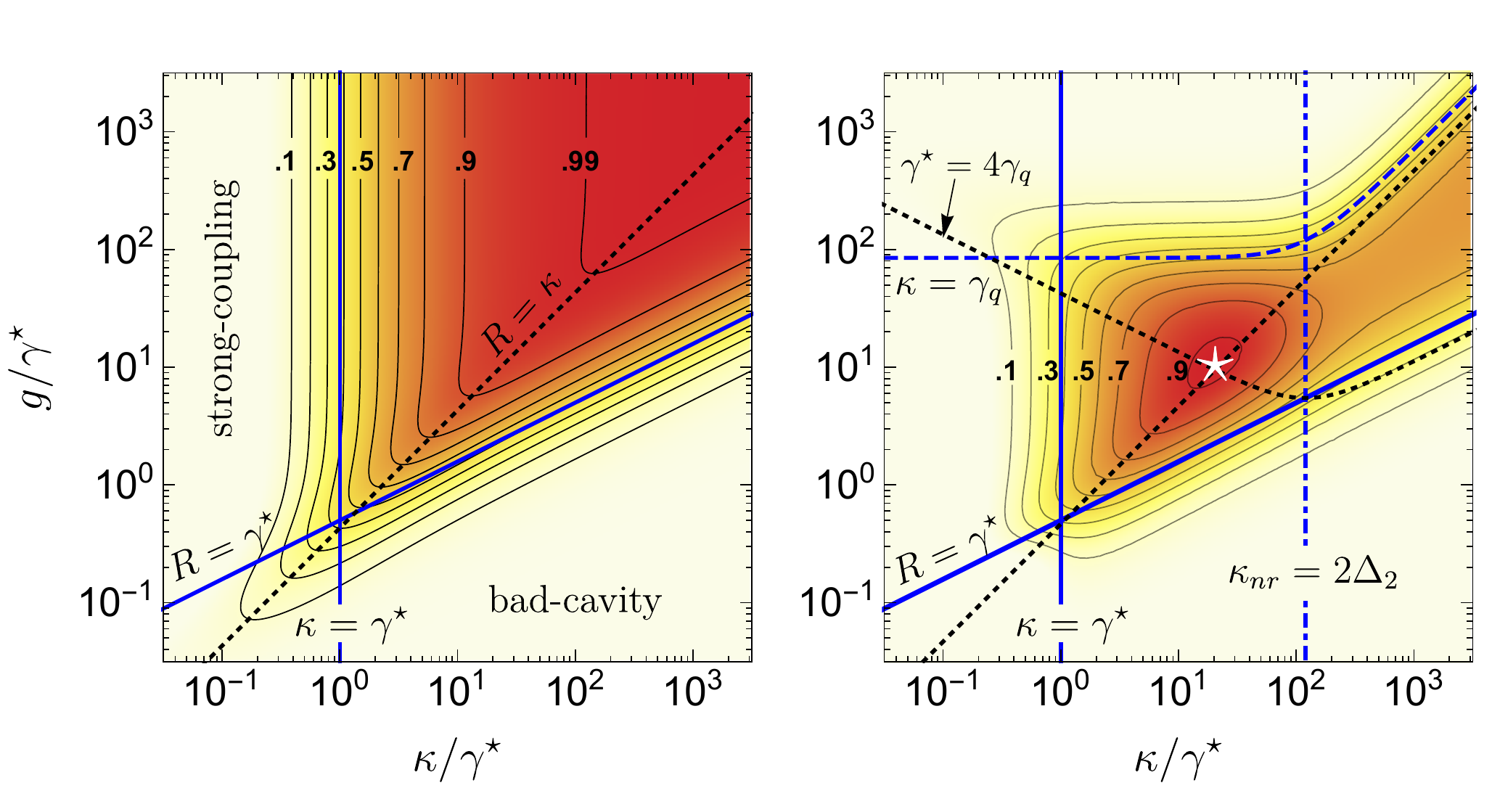}
\vspace{-7mm}
\begin{flushleft}\mbox{\hspace{4mm}(b)\hspace{36mm}(d)}\end{flushleft}
\vspace{-3mm}
\includegraphics[trim={2mm 5mm 0 0},scale=0.43]{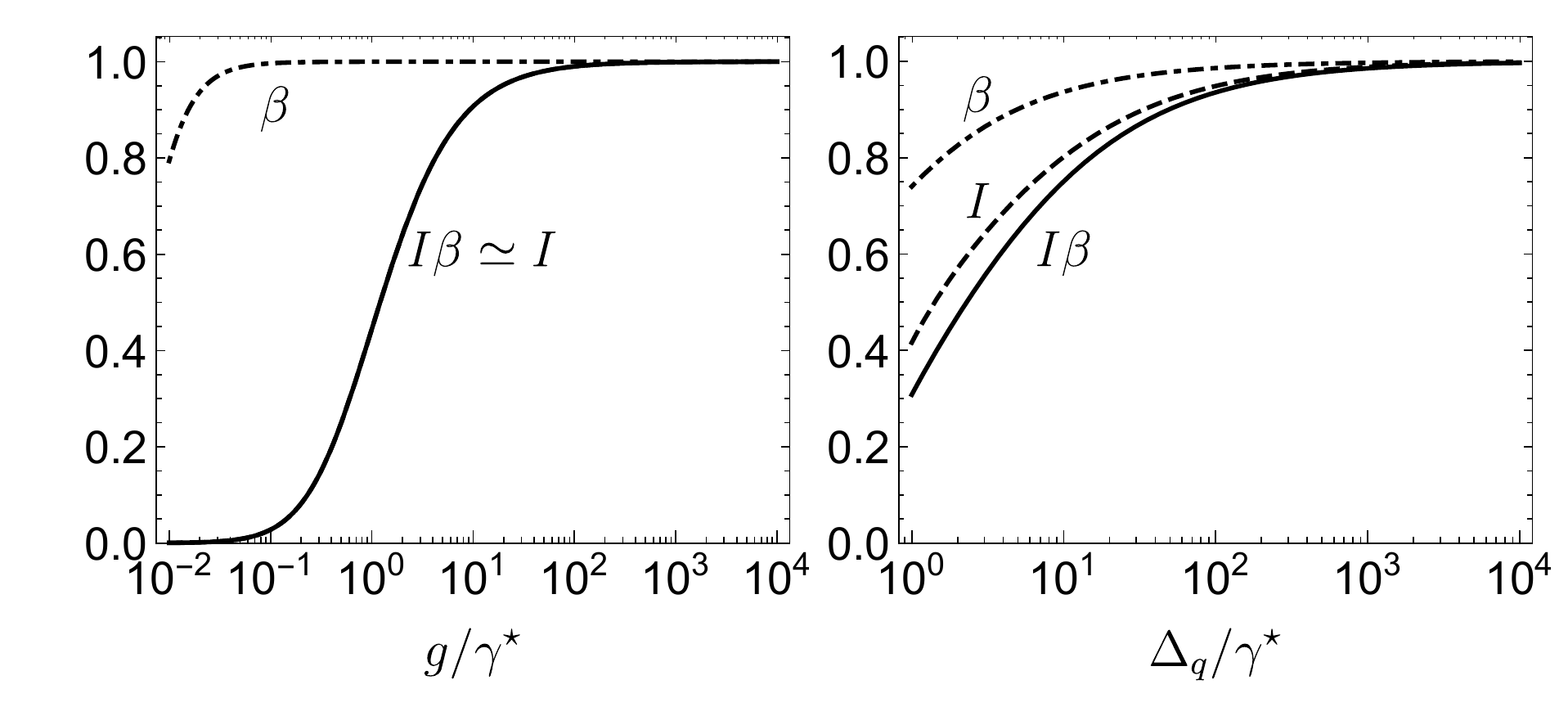}
\caption{Product of indistinguishability $I$ and intrinsic cavity efficiency $\beta$ for a Markovian emitter in the critical regime with cavity coupling $g$, emitter dephasing rate $\gamma^\star$, and cavity linewidth $\kappa$ with non-radiative portion $\kappa_\text{nr}=(1-\eta_\text{r})\kappa$ where $\eta_\text{r}$ is the bare cavity radiative efficiency. (a, b) The case without quenching ($\gamma_\text{q}$=0). (a) $I\beta$ plotted along with the critical regime boundaries (blue solid lines) and the boundary between the strong-coupling and bad-cavity regimes, $R=\kappa$ (black dotted line). (b) $I\beta$ along the $R=\kappa$ line, using $\gamma_0/\gamma^\star=10^{-4}$ to compute $\beta$. \mbox{(c, d)} The case with quenching ($\gamma_\text{q}\neq 0$). (c) $I\beta$ for a simple one-mode example, $\gamma_\text{q}=g_2^2\kappa_\text{nr}/(\Delta_2^2+(\kappa_\text{nr}/2)^2)$, to illustrate the different regimes and boundaries. Here we use $g_2=g/2$, $\Delta_2=30\gamma^\star$, and $\eta_\text{r}=0.5$, corresponding to $\Delta_\text{q}=60\gamma^\star$. The maximum of $I\beta=0.92$ occurs at the intersection between $\gamma^\star=4\gamma_\text{q}$ and $R=\kappa$ (white star). The regime above the $\kappa=\gamma_\text{q}$ line is dominated by quenching. The vertical dashed line divides the mode-detuned case to the left and the nearly resonant case to the right. (d) Maximum achievable $I\beta$ as a function of $\Delta_\text{q}/\gamma^\star$ in the mode-detuned case.}
\label{quenchingcouplingind}
\end{figure}

When there are no plasmon quenching effects and when the system is far within the critical regime boundaries, we have that $\gamma\simeq 0$ and $I_1\simeq 0$; hence, \mbox{$I=R\kappa(R+\gamma^\star)^{-1}(\kappa+\gamma^\star)^{-1}$} and $\beta=1$. From this, it can be seen that $I\beta$ is maximized when $R=\kappa$ for a given cavity coupling rate $g$ (see Fig. \ref{quenchingcouplingind} (a)). This implies $2g=\kappa$, which is also the strong-coupling boundary in the limit that $\kappa\gg\gamma+\gamma^\star$.

For a plasmonic cavity, $g$ is the coupling rate to the dominant radiating mode; however, the emitter will also couple to higher-order modes \cite{delga2014,bozhevolnyi2017,peyskens2017}. These higher-order modes contribute to the enhanced decay rate of the emitter but are predominantly non-radiative and hence quench the emission.

In our analysis, we treat higher-order modes using the Markovian approximation \cite{peyskens2017}. In this case, the emitter decay rate $\gamma$ is increased by the quenching rate $\gamma_\text{q}$ so that $\gamma$ becomes $\gamma=\gamma_\text{r}+\gamma_\text{nr}+\gamma_\text{q}$. The quenching contribution $\gamma_\text{q}$ can be described by the plasmon spectral density, which is approximated by a sum of Lorentzian functions in the quasistatic limit \cite{delga2014}:
\begin{equation}
\label{quenching}
\gamma_\text{q} = \sum_{l=2}^\infty \frac{g_l^2\kappa_\text{nr}}{\Delta_l^2+(\kappa_\text{nr}/2)^2},
\end{equation}
where $\Delta_l>0$ is the detuning of the respective mode from the emitter, $g_l=k_l g$ is the coupling rate where $k_l$ is approximately independent of $g$, and we assume that each mode has the same non-radiative rate $\kappa_\text{nr}=\kappa(1-\eta_\text{r})$ as the dominant mode \cite{peyskens2017}. This Markovian approximation is justified when the coupling rate to each individual higher-order mode is not too strong---when \mbox{$g_l^2/(\Delta_l^2+(\kappa_\text{nr}/2)^2)<1$}. Since we only require the system to achieve the optimal relation $R=\kappa$ in the critical regime, the system is described as only on the brink of strong coupling with the resonant dominant mode. Hence, we expect that the detuned higher-order modes should not display any significant strong coupling. In the critical regime, the quenching rate dominates $\gamma_\text{q}>\gamma_\text{r}+\gamma_\text{nr}$ so that $\gamma\simeq\gamma_\text{q}$.

To attain large $I\beta$, the dissipation through the cavity must be faster than the quenching rate, implying: $R> \gamma_\text{q}$ and $\kappa> \gamma_\text{q}$. There are two upper-bound cases to consider for $\gamma_\text{q}$. In the limit that the higher-order modes are near-resonant with the emitter ($4\Delta_l^2< \kappa_\text{nr}^2$), we have $\gamma_\text{q}\simeq (4g^2/\kappa_\text{nr})\sum_{l=2}^\infty k_l^2$. Then $R>\gamma_\text{q}$ implies $\sum_{l=2}^\infty k_l^2< (1-\eta_\text{r})$. This condition opposes high radiative efficiency and it is very difficult to satisfy when there are many modes, making it unsuitable. On the other hand, if the modes are detuned from the emitter \mbox{($4\Delta_l^2> \kappa_\text{nr}^2$)}, we have $\gamma_\text{q}\simeq g^2\kappa_\text{nr}/\Delta_\text{q}^2$ where we define $1/\Delta_\text{q}^2 = \sum_{l=2}^\infty k_l^2/\Delta_l^2$ for simplicity. In this case, $R,\kappa> \gamma_\text{q}$ implies that we require $\kappa,2g< 2\Delta_\text{q}(1-\eta_\text{r})^{-1/2}$ to achieve large $I\beta$. Applying the condition $\kappa>\gamma^\star$ that is required to reach the critical regime, we find the main condition: $\gamma^\star(1-\eta_\text{r})^{1/2} <2\Delta_\text{q}$.

The quantity $\Delta_\text{q}$ can be seen as an effective detuning parameter that describes the severity of quenching for a plasmonic-emitter system. The value of $\Delta_\text{q}$ depends on the geometry of the cavity and the position of the emitter relative to the cavity. For example, $\Delta_\text{q}$ for a single spherical metallic nanoparticle is dependent on the ratio $\xi=d/r$ of the distance between the emitter and the particle surface $d$ and particle radius $r$ \cite{delga2014}. For a silver sphere, $\Delta_\text{q}$ can range from $2\pi\times 6$ THz for $\xi=0.05$ to $2\pi\times 117$ THz for $\xi=2$. In this example, the limit $\gamma_\text{q}\simeq g^2\kappa_\text{nr}/\Delta_\text{q}^2$ is also a good approximation when $Q>5$.

In the absence of quenching, $I\beta$ can be increased arbitrarily by increasing both $g$ and $\kappa$ while following $R=\kappa$ to maximize the Purcell enhancement [Fig. \ref{quenchingcouplingind}]. However, in the presence of quenching, $I\beta$ decreases when $\kappa,2g>2\Delta_\text{q}(1-\eta_\text{r})^{-1/2}$. This restricts $I\beta$ to a maximum value for a given $\Delta_\text{q}$ and $\gamma^\star$. We analytically maximized $I\beta$ in the mode-detuned case for small $\gamma^\star/\Delta_\text{q}$. The values of $g$ and $\kappa$ that maximize $I\beta$ were found to be $\kappa_\text{max}\simeq 2g_\text{max}\simeq [\Delta_\text{q}^2\gamma^\star/(1-\eta_\text{r})]^{1/3}$. From this solution, we see that decreasing $\gamma^\star/\Delta_\text{q}$ or increasing $\eta_\text{r}$ will increase the maximum possible $I\beta$. We also notice that, at the $I\beta$ maximum, $\gamma_\text{q}\simeq \gamma^\star/4$ is independent of $\Delta_\text{q}$ and $\eta_\text{r}$. Hence, the maximum occurs roughly at the intersection between $R=\kappa$ and $\gamma^\star=4\gamma_\text{q}$, see Fig. \ref{quenchingcouplingind} (c).

\section{Candidate systems and phonon sideband corrections}
\label{sec:candidatePSB}

As we have shown, decreasing $\gamma^\star/\Delta_\text{q}$ can increase the maximum attainable $I\beta$. This heavily favors narrow linewidth emitters. An emitter with a smaller $\gamma^\star/\gamma_\text{r}$ ratio is also advantageous because the Purcell factor $P=R/\gamma_\text{r}$ required to reach the maximum (at $R=\kappa$ and $\gamma^\star=4\gamma_q$) is not as large. Furthermore, a good emitter for this application must have a small PSB that has a small overlap with the cavity spectrum \cite{smith2017, smith2017b}. This is necessary because any photons emitted through the PSB into the cavity will decrease the indistinguishability and any photons emitted from the PSB directly will decrease the intrinsic cavity efficiency.

The above criteria do not favor popular emitters such as quantum dots (QDs) and nitrogen-vacancy (NV$^-$) centers in diamond, both of which have significant PSBs at room temperature and generally broad ZPLs. However, a promising candidate is the negatively-charged silicon-vacancy (SiV$ ^{-}$) center in diamond. A nanodiamond SiV$ ^{-}$ center can have $\gamma^\star$ as small as $2\pi\times 380$ GHz \cite{neu2011} and $1/\gamma_\text{r}\simeq 8.3$ ns ($1/\gamma=0.58$ ns with $\gamma_\text{r}/\gamma=0.07$) \cite{benedikter2017}. They can also have a very small PSB, with a ZPL emission proportion (Debye-Waller factor) up to $\text{DW}=0.88$ \cite{neu2011}. The SiV$^-$ PSB is also spectrally well-separated from the ZPL, allowing most of the PSB emission to be filtered by the cavity.

The reduction of indistinguishability and intrinsic cavity efficiency due to PSB non-Markovian effects can be approximated in the cavity weak-coupling regime by \cite{smith2017}:
\begin{equation}
\label{Icorrect}
I = I_0\left[\frac{B^2}{B^2+F(1-B^2)}\right]^2
\end{equation}
and
\begin{equation}
\label{betacorrect}
\beta = \beta_0\frac{B^2+F(1-B^2)}{1-\beta_0(1-F)(1-B^2)},
\end{equation}
where $I_0$ and $\beta_0$ are the values computed in the Markovian approximation; $B$ is the Franck-Condon factor, $B^2=\text{DW}$, and $F$ is the fraction of the PSB not filtered by the cavity, which depends on the cavity linewidth $\kappa$ and the exact PSB spectrum for an emitter. In writing Eq. (\ref{Icorrect}), we also assume that the additional interaction between the PSB and ZPL via the cavity does not significantly alter the phonon-induced pure-dephasing rate $\gamma^\star$, which is a reasonable assumption in the regime of interest and at room temperature where $\gamma^\star$ is already quite large (see the Appendix for a more detailed discussion).

In general, $F$ increases with cavity linewidth. In the broad-cavity limit where no PSB is filtered ($F\rightarrow 1$), the $I\beta$ product is limited to $I\beta=I_0\beta_0B^4$, where $B^4$ can be interpreted as the probability that both of the photons being interfered at a beamsplitter were emitted from the ZPL. For a nanodiamond SiV$^-$ center with $\text{DW}=0.88$, this implies $I\beta\simeq 0.77I_0\beta_0$ whereas an NV$^-$ center with $\text{DW}\simeq 0.03$ \cite{santori2010} implies $I\beta\simeq 0.0009I_0\beta_0$. A cavity can improve this by allowing $F<1$, provided that the PSB is spectrally well-separated from the cavity resonance.

\begin{figure}[t]
\centering
\includegraphics[trim={6mm 5mm 0 0},scale=0.42]{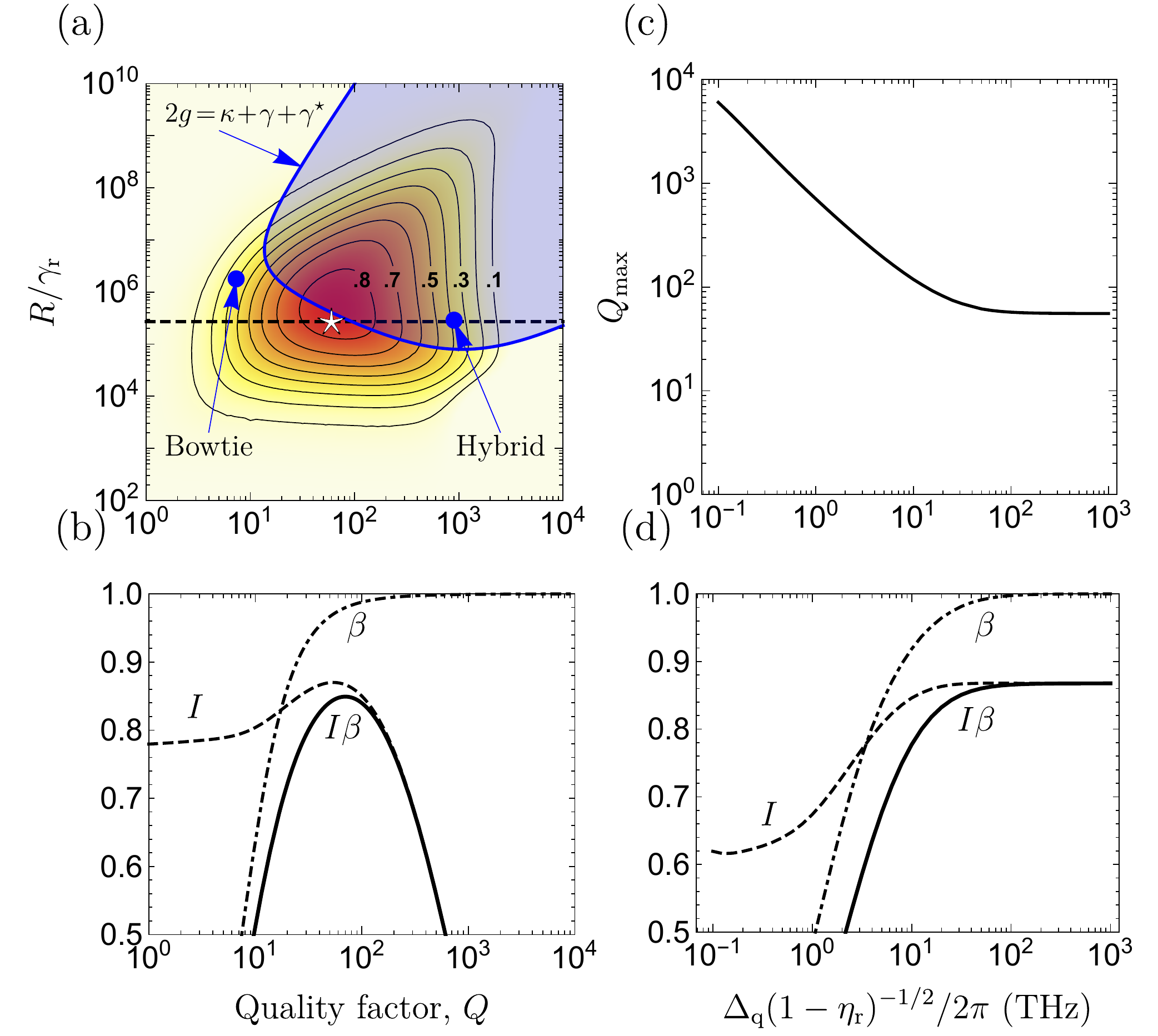}
\caption{Estimations for a nanodiamond negatively-charged silicon vacancy (SiV$^-$) center enhanced by an ultrasmall mode volume cavity when taking into account the effects of quenching and a non-zero phonon sideband (PSB). The effect of the SiV$^-$ PSB on indistinguishability $I$ and intrinsic cavity efficiency $\beta$ is estimated using an expression valid in the weak-coupling regime \cite{smith2017}. This small correction could be inaccurate in the strong-coupling regime (blue shaded region in (a)) where the zero-phonon line (ZPL) begins to display a vacuum Rabi splitting. Parameters used for the SiV$^-$ center are as follows: resonance frequency $\omega=2\pi\times 405$ THz, radiative lifetime $1/\gamma_\text{r}=8.3$ ns \cite{benedikter2017}, phonon-induced pure-dephasing rate $\gamma^\star=2\pi\times 500$ GHz \cite{neu2011}, and Debye-Waller factor $\text{DW}=0.88$ \cite{neu2011}. See the Appendix for the PSB spectrum used to calculate the correction.
(a) $I\beta$ plotted in the critical regime and in the mode-detuned case with $\Delta_\text{q}(1-\eta_\text{r})^{-1/2} = 2\pi\times 30$ THz, where $\Delta_\text{q}$ is the effective detuning parameter for higher-order non-radiative plasmon modes and $\eta_\text{r}$ is the bare cavity quantum efficiency. Here $R=4g^2/\kappa$ where $g$ is the cavity coupling rate and $\kappa$ is the bare cavity linewidth. The blue dots represent the plasmonic bowtie \cite{santhosh2016} and a plasmonic-Fabry-P\'{e}rot hybrid cavity \cite{gurlek2017}. For the bowtie, $R/\gamma_\text{r}=1.7\times 10^6$ is determined from $g=60$ meV, $Q=7.3$ \cite{santhosh2016}, and $1/\gamma_\text{r}=20$ ns \cite{brokmann2004}. The dashed line marks the $R/\gamma_\text{r}=2.7\times 10^5$ expected for the hybrid cavity (at $Q=986$) from the enhancement of the local density of states (LDOS) \cite{gurlek2017}. The white star shows the proposed single-photon source. (b) Cross-section for the dashed line in (a). The indistinguishability reaches the value of $I=\text{DW}^2\simeq 0.77$ in the limit $Q\rightarrow 0$ where the PSB is not filtered. (c) Quality factor $Q_\text{max}$ required to reach the maximum $I\beta$ for a given $\Delta_\text{q}$ and $R/\gamma_\text{r}=2.7\times 10^5$. (d) $I\beta$ expected at $Q_\text{max}$. $I$ is limited by both ZPL broadening and PSB emission in the limit that quenching is negligible.}
\label{sivplots}
\end{figure}

A promising plasmonic cavity design for room-temperature applications is the plasmonic bowtie antenna, which was used to demonstrate vacuum Rabi splitting with single QDs at room temperature \cite{santhosh2016}. Unfortunately, the close proximity of the emitter to the bowtie structure makes it difficult to achieve a large $\Delta_\text{q}$ and $\eta_\text{r}$, causing the system to be dominated by quenching. Moreover, a bowtie antenna has a very large resonance ($Q\simeq 7$), which is a poor filter for any PSB emission. These problems can be solved by placing the bowtie inside a detuned Fabry-P\'{e}rot cavity \cite{gurlek2017}. This hybrid approach promises to alleviate quenching effects, improve cavity quantum efficiency, and increase the cavity quality factor. A simulated Fabry-P\'{e}rot-bowtie hybrid cavity shows a Purcell factor (as determined from local density of states (LDOS) enhancement) up to $R/\gamma_\text{r}=2.7\times 10^5$ with a $Q$ as high as $10^3$ and near-unity radiative efficiency ($\beta\eta_\text{r}\simeq 0.95$). In addition, high collection efficiencies should be possible with these systems, e.g. 81\%\cite{gurlek2017}. Furthermore, by tuning the parameters of a hybrid cavity, it should be possible to optimize the system to produce a large $I\beta$ using the guidelines derived above. 

To estimate the $I_0\beta_0\eta_\text{r}$ achievable using a nanodiamond SiV$^-$ center inside a hybrid cavity, we used the spectrum of sample 5 from Neu \emph{et al.} \cite{neu2011} along with the predictions for cavity properties from Gurlek \emph{et al.} \cite{gurlek2017}. For the emitter, we used resonance frequency $\omega=2\pi\times 405$ THz, $1/\gamma_\text{r}=8.3$ ns \cite{benedikter2017}, and $\gamma^\star=2\pi\times500$ GHz \cite{neu2011}. For the cavity, we used $R/\gamma_\text{r}=2.7\times 10^5$, $Q=60$, and $\beta_0\eta_\text{r}=0.95$. With these parameters, the Markovian approximation gives $I_0\beta_0\eta_\text{r}=0.86$ ($I_0=0.90$, $\beta_0=0.98$) for \mbox{$\Delta_\text{q}=2\pi\times 5$ THz}. This estimate was numerically verified from Eqs. (\ref{effdef}) and (\ref{inddef}).

To approximate the effect of the PSB, we estimated the PSB spectrum using a sum of Lorentzian functions to match the measured spectrum and Debye-Waller factor $B^2=\text{DW}=0.88$ of sample 5 \cite{neu2011} (see the Appendix for the expression). For a cavity quality factor of $Q=60$, the fraction of PSB not filtered by the cavity is $F=0.15$. This leads to a correction of $I\simeq 0.96I_0$ and $\beta\simeq 0.997\beta_0$. Hence the estimation becomes $I\beta\eta_\text{r}=0.83$ ($I=0.87$, $\beta=0.97$). This correction is accurate for \mbox{$2g<\gamma+\kappa+\gamma^\star$} \cite{smith2017}. For the parameters used in this estimation we have $2g/(\gamma+\kappa+\gamma^\star)\simeq 0.8$. See Fig. \ref{sivplots} for estimations using other $Q$ and $\Delta_\text{q}$ values.

For applications in quantum information processing, it is necessary to have near-unity indistinguishability for high fidelity of quantum operations. This can be accomplished by spectrally filtering the broad emission from the hybrid cavity, which sacrifices efficiency. For an approximately-Markovian source, the $I\beta$ product after spectral filtering cannot exceed the $I\beta$ product from the source \cite{grange2015}; hence, at best it remains constant. By assuming an outcoupling of $0.81$ \cite{gurlek2017} and an initial value of $I\beta\eta_\text{r}=0.83$, this source could be capable of providing near-unity indistinguishability with a single-photon emission efficiency as high as $0.67$. This would be comparable to state-of-the-art semi-conductor sources that operate at low temperature \cite{ding2016, somaschi2016}.

Although we only discussed the SiV$^-$ center as a candidate emitter, there are other emitters that have potential. A few other diamond defects might have narrow homogeneous linewidths at room temperature, such as the N3, H2, H1b, and H1c defects \cite{stoneham2009b}. There is also evidence that the infrared transition of the NV$^-$ center is weakly coupled to phonons \cite{rogers2008, doherty2014, rogers2015} and could have a very narrow homogeneous linewidth with a relatively small spectrally-separated PSB \cite{rogers2008}. In addition, the neutral silicon vacancy (SiV$^0$) center shows a promising combination of optical and spin properties \cite{green2017, rose2017}. Its symmetry and electronic configuration allows it to exhibit stable optical properties and long spin coherence times, combining the best aspects of the SiV$^-$ and NV$^-$ centers, respectively. Moreover, it has a reported Debye-Waller factor of $\text{DW}>0.9$---a single-defect measurement limited only by the noise floor \cite{rose2017}. If the spin properties of the SiV$^0$ center can be made comparable to those of the NV$^-$ center, it could function (in combination with a hybrid plasmonic cavity) as a spin-photon interface for a room-temperature solid-state quantum network.

An interesting alternative approach could be to use proposed pure-dielectric ultrasmall cavities \cite{robinson2005,hu2016,choi2017}. These cavities should maintain high quality factors ($Q\simeq 10^6$) with mode volumes as small as $7\times 10^{-5}\lambda^3$ providing unprecedented Purcell factors without being affected by plasmon quenching \cite{choi2017}. These high-Q ultrasmall cavities would need to be used with very narrow emitters, since the cavity must decay faster than the pure-dephasing rate ($\kappa>\gamma^\star$). In particular, single rare-earth ions are known for having very narrow but dim lines \cite{perrot2013}. These quantum emitters can contain very phonon-resistant transitions (such as the $^5$D$_0\rightarrow ^7$F$_0$ transition in the europium(III) ion \cite{binnemans2015}) that might remain quite narrow at room temperature and exhibit a very small phonon sideband. Such a pure-dielectric ultrasmall cavity could also be useful with broader emitters if its cavity quality factor could be lowered to \mbox{$\sim 10^2$--$10^3$} without sacrificing an increase in the effective mode volume.

\vspace{-3mm}
\section{Conclusion}
\vspace{-3mm}
We have shown that highly efficient solid-state room-temperature indistinguishable SPSs should be within reach using ultrasmall mode volume cavities, and we have described the most promising regime of operation. For cavities containing plasmonic materials, this regime exists only for emitters that are narrow and bright enough to allow a large effective Purcell enhancement without being dominated by plasmon quenching. In addition, it exists only for emitters with a small PSB that is spectrally well-separated from its ZPL. In particular, a nanodiamond SiV$^-$ defect combined with a hybrid plasmonic-Fabry-P\'{e}rot cavity appears to be exceptionally promising. Room-temperature indistinguishable SPSs would be a significant advance for quantum technology, while also helping to answer fundamental questions about the physical regimes in which quantum phenomena can be observed \cite{plumhof2014, kumar2016}.

\vspace{-3mm}
\section*{Acknowledgments}
\vspace{-3mm}
This work was supported by the Natural Sciences and Engineering Research Council of Canada (NSERC) Discovery Grant (DG), Canadian Graduate Scholarships-Master's (CGSM), and Undergraduate Student Research Awards (USRA) programs; the Program for Undergraduate Research Experience (PURE) and Eyes High program at the University of Calgary; and Alberta Innovates Technology Futures (AITF) Graduate Student Scholarships (GSS). The authors thank Paul Barclay, Gaston Hornecker, Thomas Lutz, Tamiko Masuda, Neil Sinclair, and Wolfgang Tittel for useful discussions.

\appendix
\setcounter{equation}{0}
\renewcommand{\theequation}{S\arabic{equation}}
\section*{Appendix}
\label{appendix}

\subsection{Optical Bloch Equations}
\label{app:blocheq}
To derive the indistinguishability, we begin by writing the optical Bloch equations in the single-excitation regime given the Hamiltonian and master equation (Eqs. (\ref{hamiltonian}) and (\ref{mastereq}) in the main text). The optical Bloch equations are:
\begin{equation}
\frac{\text{d}}{\text{d}t}\left[\begin{matrix}
\braket{\hat{a}^\dagger}\\
\braket{\hat{\sigma}_+}
\end{matrix}\right]
=
A_1
\left[\begin{matrix}
\braket{\hat{a}^\dagger}\\
\braket{\hat{\sigma}_+}
\end{matrix}\right],
\end{equation}
\begin{equation}
\frac{\text{d}}{\text{d}t}\left[\begin{matrix}
\braket{\hat{a}^\dagger\hat{a}}\\
\braket{\hat{a}^\dagger\hat{\sigma}_-}\\
\braket{\hat{\sigma}_+\hat{a}}\\
\braket{\hat{\sigma}_+\hat{\sigma}_-}\\
\end{matrix}\right]
=
A_2
\left[\begin{matrix}
\braket{\hat{a}^\dagger\hat{a}}\\
\braket{\hat{a}^\dagger\hat{\sigma}_-}\\
\braket{\hat{\sigma}_+\hat{a}}\\
\braket{\hat{\sigma}_+\hat{\sigma}}\\
\end{matrix}\right],
\end{equation}
where
\begin{equation}
\label{A1}
A_1=-\frac{1}{2}\left[\begin{matrix}
\kappa&-2ig\\
-2ig&\gamma^\star+\gamma
\end{matrix}\right],
\end{equation}
\begin{equation}
\label{A2}
A_2=-\frac{1}{2}\left[\begin{matrix}
2\kappa&2ig&-2ig&0\\
2ig&\kappa+\gamma^\star+\gamma&0&-2ig\\
-2ig&0&\kappa+\gamma^\star+\gamma&2ig\\
0&-2ig&2ig&2\gamma\\
\end{matrix}\right].
\end{equation}

\subsection{Efficiency}
\label{ssec:efficiency}
The cavity efficiency is defined by $\beta = \kappa\int_0^\infty \braket{\hat{a}^\dagger(t)\hat{a}(t)}\text{d}t$ in the main text. An exact solution to $\beta$ can be derived by integrating the solution to Eq. (\ref{A2}). Assuming instantaneous excitation of the system, we set the initial condition to $\rho(0)=\ket{e}\bra{e}$. This implies that the only nonzero initial condition for the optical Bloch equations is $\braket{\hat{\sigma}_+(0)\hat{\sigma}_-(0)}=1$. With this initial condition, we have:
\begin{equation}
\begin{aligned}
\beta&=\kappa\int_0^\infty\text{exp}\!\left(tA_2\right)_{14}\text{d}t = -\kappa \left(A_2^{-1}\right)_{14}\\
&=\frac{4g^2\kappa}{4g^2(\gamma+\kappa)+\gamma\kappa(\gamma+\kappa+\gamma^\star)},
\end{aligned}
\end{equation}
which holds in all parameter regimes for a Markovian system and is equal to the result given in \cite{grange2015}. By substituting \mbox{$g=(R\kappa)^{1/2}/2$} into this expression, it can be arranged to the form given in the main text. Also, notice that $\gamma\ll\kappa,g,\gamma^\star$ gives $\beta=1$.

\subsection{Indistinguishability}
\label{ssec:indistinguishability}
We can analytically solve for the indistinguishability in a way similar to that for the efficiency. We are interested in the regime where $\gamma^\star$ is small relative to $g$ and $\kappa$. Since $A_1$ can be easily diagonalized, the evolution given by $A_1$ can be solved exactly by computing $U(t)=\text{exp}\!\left(A_1t\right)$. However, to simplify solving the propagator $W(t)=\text{exp}\!\left(A_2t\right)$, we treat it perturbatively for $\gamma^\star/(\kappa+\gamma)<1$. Interestingly, this condition is satisfied if we only assume $\gamma^\star<\kappa$ and so we need not make assumptions about the relative magnitude of $\gamma$ and $\gamma^\star$ or $g$ and $\gamma^\star$.

Let $A_2^{(0)}=A_2(\gamma^\star=0)$, and let $A_2^{(1)}=A_2-A_2^{(0)}$. Then we can write $W(t)= W^{(0)}+W^{(1)}+\mathcal{O}\left(\gamma^{\star 2}\right)$ where:
\begin{equation}
\begin{aligned}
W^{(0)}&=\text{exp}\!\left(A_2^{(0)}t\right),\\
W^{(1)}&=W^{(0)}\int_0^t \text{exp}\!\left(-A_2^{(0)}t^\prime\right)A_2^{(1)}\text{exp}\!\left(A_2^{(0)}t^\prime\right)\text{d}t^\prime.
\end{aligned}
\end{equation}
The indistinguishability is defined by Eq. (\ref{inddef}) in the main text. Since the Hamiltonian is time independent, this expression can be simplified to:
\begin{equation}
I = \frac{2\kappa^2}{\beta^2}\int_0^\infty\int_0^\infty|\braket{\hat{a}^\dagger(t+\tau)\hat{a}(t)}|^2\text{d}t\text{d}\tau.
\end{equation}
From the quantum regression theorem \cite{gardiner2004}, we can write:
\begin{equation}
\braket{\hat{a}^\dagger(t+\tau)\hat{a}(t)} = U_{11}(\tau)\braket{\hat{a}^\dagger(t)\hat{a}(t)}+U_{12}(\tau)\braket{\hat{\sigma}_+(t)\hat{a}(t)},
\end{equation}
where the subscripts denote the element of the matrix propagator $U$. Taking the initial condition to be $\braket{\hat{\sigma}_+(0)\hat{\sigma}_-(0)}=1$, we have $\braket{\hat{a}^\dagger(t)\hat{a}(t)} = W_{14}$ and $\braket{\hat{\sigma}_+(t)\hat{a}(t)} = W_{34}$. Then the correlation function becomes:
\begin{equation}
\braket{\hat{a}^\dagger(t+\tau)\hat{a}(t)} = U_{11}(\tau)W_{14}(t)+U_{12}(\tau)W_{34}(t).
\end{equation}
From this, we can write the indistinguishability as:
\begin{equation}
\begin{aligned}
I\beta^2 &= 2\kappa^2\left[C_1+2\text{Re}(C_2)+C_3
\right],
\end{aligned}
\end{equation}
where
\begin{equation}
\begin{aligned}
C_1&=\int_0^\infty|U_{11}|^2\text{d}\tau\int_0^\infty|W_{14}|^2\text{d}t,\\
C_2&=\int_0^\infty U^\star_{11}U_{12}\text{d}\tau\int_0^\infty W^\star_{14}W_{34}\text{d}t,\\
C_3&=\int_0^\infty|U_{12}|^2\text{d}\tau\int_0^\infty|W_{34}|^2\text{d}t.
\end{aligned}
\end{equation}

This expression can be split into two parts: $I=I^{(0)}+I^{(1)}+\mathcal{O}({\gamma^\star}^2/(\kappa+\gamma)^2)$ where $I^{(0)}$ is computed using terms with $U$ and $W^{(0)}$. After much algebra and calculus, we found $I^{(0)}$ to be:
\begin{equation}
I^{(0)}\beta^2=\frac{R^2\kappa ^2 \left[3\gamma^\star(2\gamma+3\kappa+\gamma^\star)+\Gamma_1^2\right]}{(R+\gamma)(\kappa+\gamma)(R+\gamma+\gamma^\star)(\kappa+\gamma+\gamma^\star) \Gamma_1^2},
\end{equation}
where $\Gamma_1^2=(3 \gamma +\kappa ) (\gamma +3 \kappa )+4 \kappa  R$.

The perturbation $I^{(1)}$ contains the correction required for $I$ to be exact to first-order in $\gamma^\star/(\kappa+\gamma)$. Since we are only interested in the first-order correction in the perturbation of $W(t)$, we only compute the $W$ cross-terms that are first-order in $\gamma^\star/(\kappa+\gamma)$. This means, for example, only computing terms with ${W^{(0)}}^\star_{14}W^{(1)}_{34}\propto\gamma^\star$ but not those with ${W^{(1)}}^\star_{14}W^{(1)}_{34}\propto{\gamma^\star}^2$. However, we keep $U(t)$ terms exact because any expansion of $U(t)$ in $\gamma^\star$ would require the assumption that $\gamma^\star/\gamma<1$, which is not the desired case. For this reason, the result for $I^{(1)}$ for arbitrary $\gamma$ still contains some higher-order $\gamma^\star$ terms:

\begin{equation}
\begin{aligned}
\frac{I^{(1)}}{\gamma^\star I^{(0)}}&=\frac{(R-2 \gamma ) \left[(\gamma +\gamma^\star )^2+\gamma  \kappa \right]}{ \left[3\gamma^\star(2\gamma+3\kappa+\gamma^\star)+\Gamma_1^2\right]\Gamma_2^2}\\
   &-\frac{\gamma^\star(\gamma-\gamma^\star)  (4 \gamma +R)+2 \gamma  (\gamma +R) (2 \gamma +R)}{2 (\gamma +\kappa ) (\gamma +R) \Gamma_2^2}\\
    &-\frac{(\gamma +\kappa ) (8 \gamma +5 R)}{2 (\gamma +R) \Gamma_1^2},
 \end{aligned}
\end{equation}
where $\Gamma_2^2=3\gamma^\star(\gamma -\gamma^\star)+4 \gamma  (\gamma +R)$.

We use this full solution to compute estimations and generate plots in the main text. However, the expression can be simplified under the assumption that $\gamma\ll\gamma^\star$, which would arise when quenching is weak. In this case, the expression $I=I^{(0)}+I^{(1)}$ can be reduced to the form given by Eq. (\ref{ind}) in the main text.

\subsection{Phonon sideband corrections}
\label{ssec:psb}

To estimate the correction to $I\beta$ expected due to the presence of a phonon sideband (PSB) in the SiV$^-$ spectrum, we first estimated the fraction $F$ of PSB not removed by the cavity \cite{smith2017}. In terms of wavelength, this can be defined as:
\begin{equation}
F(Q)=\frac{\int_0^\infty S_\text{cav}(\lambda,Q)\times S_\text{PSB}(\lambda)\text{d}\lambda}{\int_0^\infty S_\text{PSB}(\lambda)\text{d}\lambda},
\end{equation}
where
\begin{equation}
S_\text{cav}(\lambda,Q)=\frac{1}{1+4Q^2\frac{(\lambda-\lambda_0)^2}{\lambda_0^2}},
\end{equation}
and
\begin{equation}
S_\text{PSB}(\lambda) = \sum_i\frac{a_i}{1+\frac{(\lambda-\lambda_0-c_i)^2}{b_i^2}}.
\end{equation}
The coefficients $(a_i,b_i,c_i)$ were determined by visually fitting the total spectrum $S(\lambda) = S_\text{ZPL}(\lambda) + S_\text{PSB}(\lambda)$ to the measured spectrum of samples reported by Neu \emph{et al.} \cite{neu2011}, while also maintaining that:
\begin{equation}
\text{DW} \simeq \frac{\int_0^\infty S_\text{ZPL}(\lambda)\text{d}\lambda}{\int_0^\infty S(\lambda)\text{d}\lambda},
\end{equation}
for the associated Debye-Waller ($\text{DW}$) values reported for that sample. Here we use:
\begin{equation}
S_\text{ZPL}(\lambda) = \frac{1}{1+\frac{(\lambda-\lambda_0)^2}{\delta^2}},
\end{equation}
where $\lambda_0$ is the zero-phonon line (ZPL) resonance of the sample and $\delta$ is the ZPL width in wavelength. We chose to fit the sample with the smallest $\delta$ (sample 3), and the sample with the largest $\text{DW}$ (sample 5). The PSB coefficients that we determined are given in Table \ref{coefTable}.

\begin{figure}[t!]
\includegraphics[trim={6mm 5mm 0 0},scale=0.41]{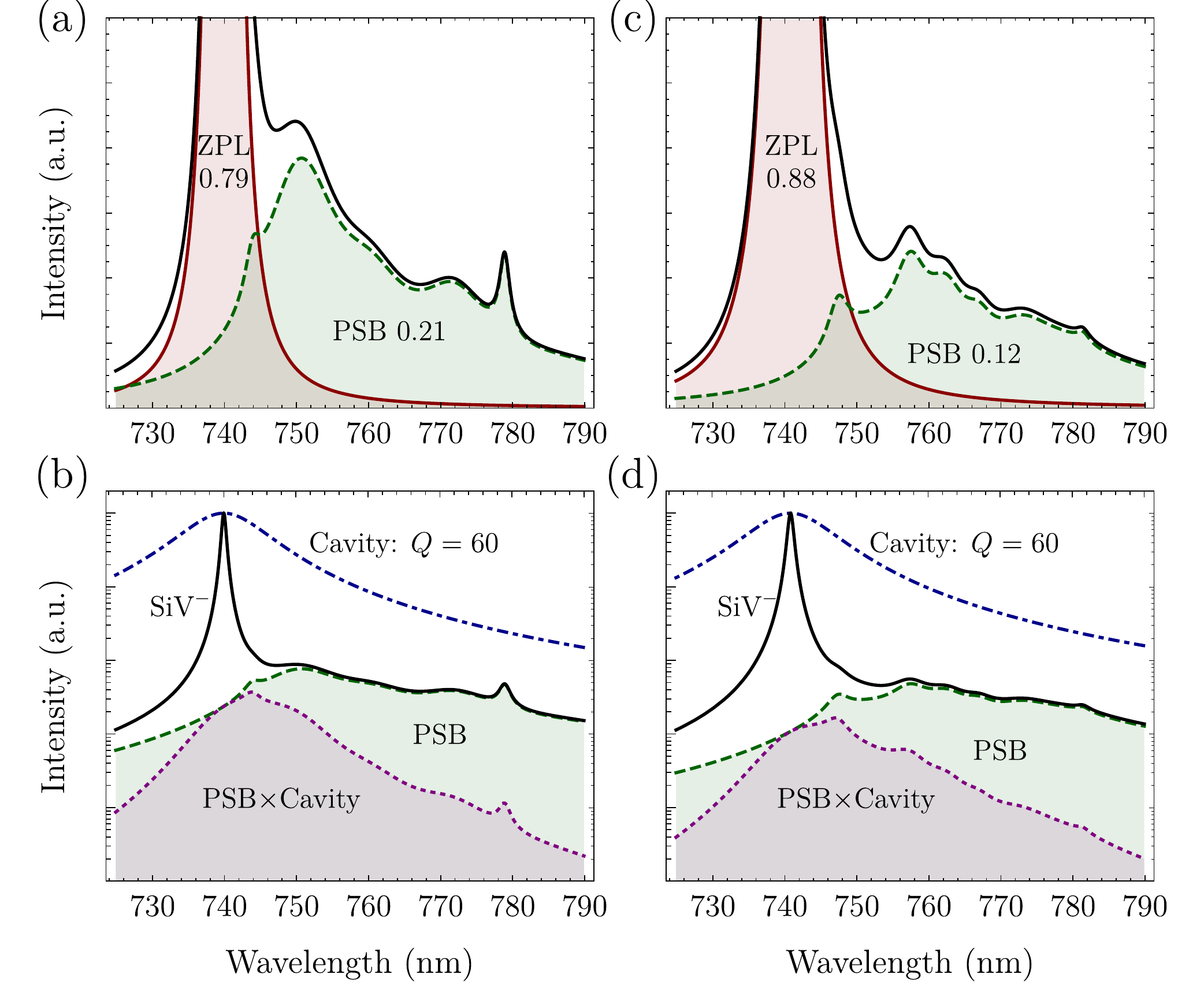}
\vspace{-2mm}
\caption{Estimated nanodiamond negatively-charged silicon-vacancy (SiV$^-$) center emission spectrum of sample 3 (a,b) and sample 5 (c,d) from Neu $\emph{et al.}$ \cite{neu2011}. (b,d) Intensity log-scale plot illustrating the reduction of the PSB due to a cavity with quality factor $Q=60$ on resonance with the zero-phonon line (ZPL) of the sample.}
\label{spectrumplots}
\end{figure}

The estimated spectra for samples 3 and 5 are illustrated in Fig. \ref{spectrumplots}. The corresponding DW factors were calculated to be $\text{DW}_3=0.791$ and $\text{DW}_5=0.884$, which closely match the measured values of $0.79$ and $0.88$ for samples 3 and 5, respectively. For a cavity with $Q=60$, we estimated the fraction $F$ to be $F_3(60) = 0.19$ and $F_5(60)=0.15$. Using $B^2=\text{DW}$ and the estimated values for $F(Q)$, we applied the PSB corrections using equations (\ref{Icorrect}) and (\ref{betacorrect}) to our results for indistinguishability $I_0$ and intrinsic cavity efficiency $\beta_0$ under the Markovian approximation.

\begin{table}[t]
\caption{Coefficients for $S_\text{PSB}(\lambda)$ used to represent the nanodiamond negatively-charged silicon vacancy center spectrum of samples 3 and 5 from Neu \emph{et al.} \cite{neu2011}.}
\begin{center}
\mbox{\hspace{5mm}Sample 3\hspace{23mm} Sample 5}
\vspace{-6.5mm}
\end{center}
\begin{tabular}{c|ccc|ccc}
 & & & & & & \\
i&$a_i\times 10^{3}$&$b_i$ (nm)&$c_i$ (nm) &$a_i\times 10^{3}$&$b_i$ (nm)&$c_i$ (nm) 
\\\hline
1&1.4        &1.3     &4.0      &1.4             &1.7     &7.5\\
2&6.7        &6.5     &10.5     &1.8             &8.0     &11.0\\
3&2.0        &6.0     &20.5     &2.6             &2.9     &17.5\\
4&2.2        &6.0     &32.0     &2.0             &3.3     &22.5\\
5&2.4        &0.9     &39.0     &0.9             &2.5     &27.0\\
6&1.0        &20.0    &47.0     &1.2             &5.0     &33.0\\
7&           &        &         &1.0             &8.0     &39.0\\
8&           &        &         &0.3             &1.1     &41.5\\
9&           &        &         &0.7             &18.0    &49.0\\
\end{tabular}
\label{coefTable}
\end{table}

The narrower linewidth of sample 3 would provide a higher $I_0\beta_0$ than sample 5. However, the larger PSB of sample 3 would restrict the indistinguishability for lower $Q$ values. Hence the maximum $I\beta$ for sample 3 is attained at $Q=100$ rather than $Q=60$. For $Q=100$, $R/\gamma_\text{r}=2.7\times 10^5$, and $\Delta_q(1-\eta_\text{r})^{-1/2}=2\pi\times 30$ THz, sample 3 reaches $I=0.85$ and $\beta=0.99$ giving $I\beta=0.84$, which is comparable to sample 5 at $Q=60$ ($I=0.87$, $\beta=0.97$, $I\beta=0.84$).

\begin{figure}[h!]
\vspace{-4mm}
\centering
\includegraphics[trim={0mm 12mm 0 0},scale=0.42]{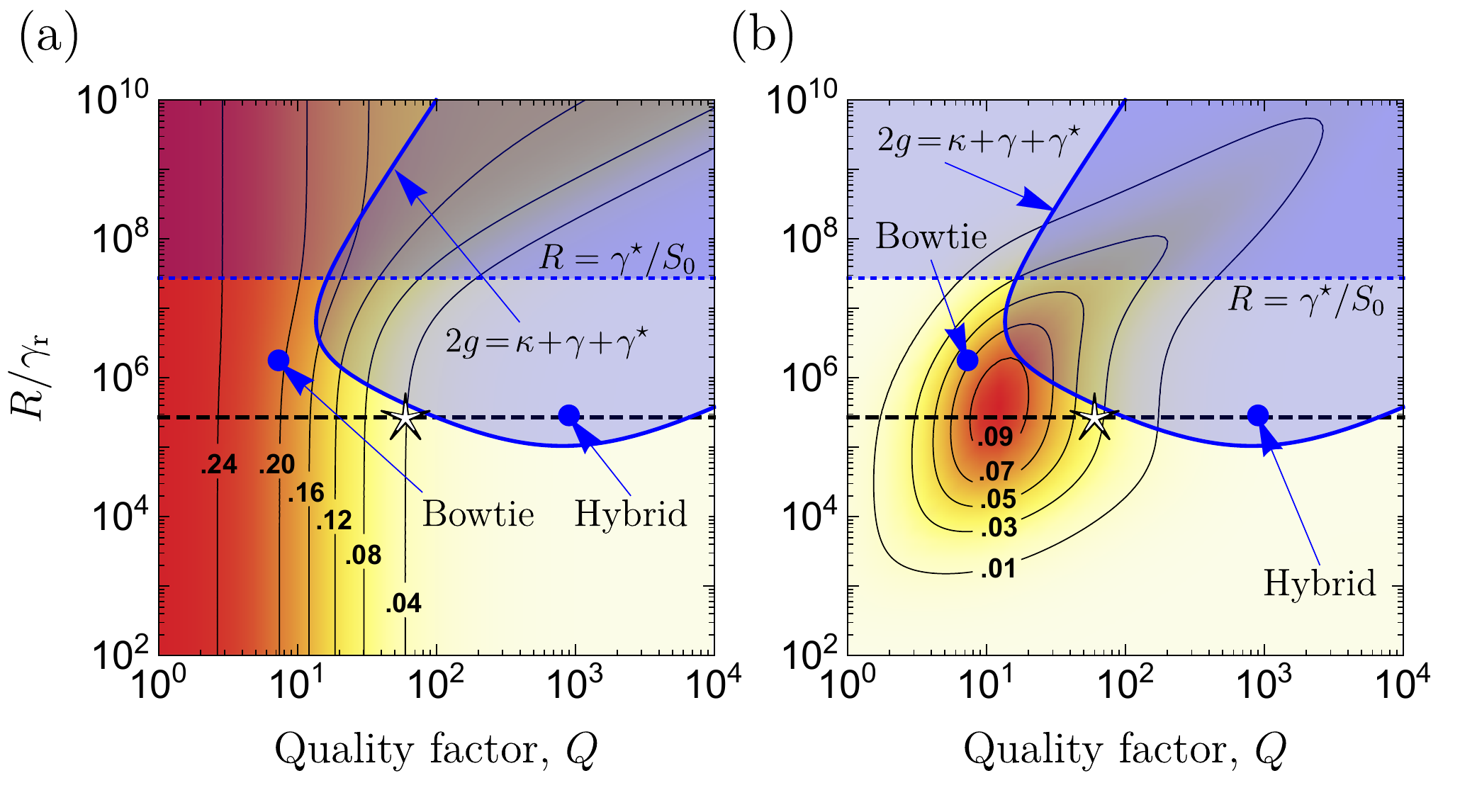}
\caption{(a) The relative error $2(I_0\beta_0-I\beta)/(I_0\beta_0+I\beta)$ for the PSB correction plotted in the critical regime and in the mode-detuned case with $\Delta_\text{q}(1-\eta_\text{r})^{-1/2} = 2\pi\times 30$ THz, where $\Delta_\text{q}$ is the effective detuning parameter for higher-order non-radiative plasmon modes and $\eta_\text{r}$ is the bare cavity quantum efficiency. Here $R=4g^2/\kappa$ where $g$ is the cavity coupling rate and $\kappa$ is the bare cavity linewidth. $I_0\beta_0$ is the derived estimation under the Markovian approximation and $I\beta$ is the value after including the correction due to the phonon sideband (PSB). This small correction could be inaccurate in the strong-coupling regime ($2g>\kappa+\gamma^\star+\gamma$) and when the PSB might begin to enhance the ZPL dephasing rate $R>\gamma^\star/S_0$ (blue shaded regions). Parameters used for the SiV$^-$: $\omega=2\pi\times 405$ THz, $1/\gamma_\text{r}=8.3$ ns \cite{benedikter2017}, $\gamma^\star=2\pi\times 500$ GHz \cite{neu2011}, $\text{DW}=0.88$ \cite{neu2011}, and $S_0=9.6\times 10^{-4}$. See the Appendix text for the PSB spectrum used to calculate the correction. The blue dots represent the plasmonic bowtie \cite{santhosh2016} and a plasmonic-Fabry-P\'{e}rot hybrid cavity \cite{gurlek2017}. For the bowtie, a $R/\gamma_\text{r}=1.7\times 10^6$ ratio is determined from $g=60$ meV, $Q=7.3$ \cite{santhosh2016}, and $1/\gamma_\text{r}=20$ ns \cite{brokmann2004}. The dashed line marks $R/\gamma_\text{r}=2.7\times 10^5$ expected for the hybrid cavity (at $Q=986$) from the enhancement of the local density of states (LDOS) \cite{gurlek2017}. The white star shows the proposed single-photon source. (b) Absolute difference $I_0\beta_0-I\beta$ plotted to illustrate the magnitude of the correction.}
\label{errorplots}
\vspace{-6mm}
\end{figure}

The regime in which the Markovian approximation is valid depends strongly on the shape and size of the emitter's PSB. In general, the approximation becomes less accurate as the cavity quality factor is decreased (see Fig. \ref{errorplots} (a)). In addition, when computing $I_0$ to use in equation (\ref{Icorrect}), we assume that the phonon-induced pure-dephasing rate is not significantly altered by the coupling between the PSB and the ZPL via the cavity mode. A sufficient condition to ensure that this assumption is valid is given when the pure-dephasing rate is much larger than the total effective rate between the PSB and the cavity mode: $\gamma^\star\gg R S_\text{PSB}(\lambda_0)$. In this case, the cavity-PSB interaction can be considered as predominantly a filtering effect. For samples 3 and 5, we can estimate $S_0=S_\text{PSB}(\lambda_0)$ as $2.4\times 10^{-3}$ and $9.6\times 10^{-4}$, respectively. For $R>\gamma^\star/S_0$, it may be necessary to consider the influence of the cavity-PSB interaction on the ZPL dephasing rate.

We also note that we have not considered interactions between the PSB and the higher-order plasmon modes. However, the PSB is predominantly Stokes-shifted to lower energy whereas the higher-order plasmon modes are generally of higher energy than the cavity resonance. Hence, the higher-order plasmon modes should not enhance the PSB and so this additional interaction can be safely neglected in the same regime where quenching does not dominate.

For narrow emitters with a spectrally-separated PSB, such as the nanodiamond SiV$^-$, the $I\beta$ maximum can exist in a parameter regime where the detrimental effects of the PSB are small, leaving the maximum to be primarily limited by Markovian processes (see Fig.\ref{errorplots} (b)). This leaves a cavity-parameter range where efficient indistinguishable emission should be attainable at room temperature.

\newpage

\bibliography{refs}

\end{document}